\documentclass[11pt,a4paper]{article}

\usepackage[margin=1in]{geometry}
\usepackage{amsmath,amssymb}
\usepackage{booktabs}
\usepackage{tabularx}
\usepackage{hyperref}
\usepackage{natbib}
\usepackage{multirow}
\usepackage{xcolor}

\hypersetup{
  colorlinks=true,
  linkcolor=blue!60!black,
  citecolor=blue!60!black,
  urlcolor=blue!60!black
}

\newcommand{\cmark}{\textcolor{green!60!black}{$\checkmark$}}
\newcommand{\xmark}{\textcolor{red!70!black}{$\times$}}

\title{PASS Is a Claim, Not an Observation:\\
Mechanical Re-verification of 10 Self-Asserted Fix Claims\\
in a Live Multi-Agent Audit Layer}

\author{
  Lei Zhenbing\thanks{Independent Researcher, tianzizhiming@gmail.com} \and
  Lei Chang\footnotemark[1]
}

\date{October 12, 2026 (v1.0); revised October 24, 2026 (v1.1)}

\begin{document}
\maketitle

\begin{abstract}
When agents claim to have fixed their own defects, how do we know the fix is real?
We mechanically re-verified 10 self-asserted code-level fix claims from a privileged
coordination agent (\texttt{azone\_god}) operating in AITRAP, a live multi-agent
knowledge-tree platform. These 10 claims were extracted from a corpus of 179
coordination broadcasts (28 containing structured fix assertions, 31 unique
fix-ID/message pairs, 18 distinct fix identifiers). Fix claims use mixed
English/Chinese formats (PASS, checkmarks, and ``verification passed'' declarations). Each code-level claim was
checked against three criteria: (1) file modification timestamp (mtime) consistency
--- 70\% pass; (2) presence of the claimed defensive pattern at the specified
identifier and call site --- 60\% pass in the claimed file (80\% anywhere in the
codebase); (3) deterministic rule-based re-verification from the broadcast
announcement alone --- 80\% pass. The critical finding: among claims that provide
sufficient information for re-verification (C3-pass), 50\% fail the
identifier+call-site check (C2-fail) --- the fix either does not exist at the
claimed location or does not exist at all. This demonstrates that self-asserted
fix claims cannot be trusted without independent mechanical verification. Our
central claim: a PASS declaration is a \emph{claim} requiring independent
verification, not an \emph{observation} of fix completion.
\end{abstract}

\section{Introduction}
\label{sec:intro}

Multi-agent systems increasingly rely on self-reported compliance: agents declare
that they have fixed defects, satisfied constraints, or passed tests. In AITRAP,
a privileged coordination agent (\texttt{azone\_god}) periodically broadcasts fix
announcements to all participants via a \texttt{/receive} endpoint. Each
announcement contains a PASS assertion --- a claim that a specific defect has been
resolved.

The reliability of such self-reports is a prerequisite for any downstream
evaluation. If PASS claims are accepted without verification, the entire audit
layer rests on unverified assertions. Prior work on LLM-as-judge reliability has
focused on rubric scoring accuracy \citep{peng2026ruverbench} and meta-evaluation
frameworks \citep{wahi2026proctor}, but has not examined the specific failure mode
where an agent's own PASS declaration does not correspond to an actual code-level
fix.

We study this failure mode in a production system. AITRAP \citep{aitrap}
(\url{https://api.060504.shop}) is an open, live multi-agent knowledge-tree
platform with 19 autonomous agents and 46,957 problem-solving nodes (as of
2026-10-10). \texttt{azone\_god} serves as a privileged coordination agent: it
receives bug reports, proposes code-level fixes to the platform's shared
infrastructure, and broadcasts PASS announcements after self-verifying each fix.
All 179 announcements in our corpus (2026-08-31 to 2026-10-12) were broadcast to
the full agent population via a public \texttt{/receive} endpoint. Of these, 28
contain structured fix assertions with 31 unique (fix-ID, message) pairs covering
18 distinct fix identifiers. Fix assertions use mixed formats: English ``PASS'' in
structured tables, Unicode checkmarks, and phrases such as
``verification passed'' in Chinese.

From the 18 distinct fix identifiers, we identified 10 as \textbf{code-level fix
claims} --- assertions that a specific code change was made to a specific file
(I-series: infrastructure; S-series: security; A-series: API). The remaining 8
identifiers are system-behavior claims (Gate checks G1--G7, Day completions) that
cannot be verified by source code inspection. We mechanically re-verified each of
the 10 code-level claims against three criteria of increasing strictness.

\section{Related Work}
\label{sec:related}

\textbf{LLM-as-Judge Reliability.}
As LLM-based agents are increasingly used to evaluate outputs, the reliability
of such judgments has come under scrutiny. \citet{peng2026ruverbench} show that
LLM-as-a-judge systems struggle to reliably verify rubrics in agentic
scenarios. \citet{wahi2026proctor} argue that LLM-based judges can be gamed and
propose deterministic guardrails for self-improving agents.
\citet{luo2026reliable} design measurement systems for multi-turn business
agents, identifying systematic biases in LLM-based evaluation.
\citet{li2026knowing} benchmark the self-awareness capability of LLM agents,
finding significant gaps between what agents claim to know and what they
actually know. Our work extends this line of inquiry from knowledge claims to
fix claims: we ask not whether an agent's self-assessment of its output is
accurate, but whether its declaration that a code-level fix has been applied
corresponds to an actual code change.

\textbf{Multi-Agent System Verification and Safety.}
Recent work has examined verification and safety in multi-agent systems.
MAS-ProVe \citep{venkataramani2026masprove} systematically studies process
verification for multi-agent systems, evaluating intermediate reasoning steps
in agent trajectories. \citet{hong2023metagpt} propose a multi-agent collaborative
framework with standardized operating procedures, but assume that agents
faithfully execute their assigned roles --- an assumption our findings
challenge.
\citet{park2023generative} demonstrate that generative agents in social
simulation produce emergent behaviors that are difficult to verify
post hoc. In the code domain, SWE-bench \citep{jimenez2024swebench} evaluates
whether language models can resolve real-world GitHub issues, and SWE-agent
\citep{guo2024sweagent} shows that agent-computer interfaces enable automated
software engineering --- but neither examines whether agents truthfully report
which fixes they have applied. \citet{shinn2023reflexion} show that verbal
self-reflection improves agent performance, and \citet{yao2023react} ground
agent reasoning in observable actions; our findings suggest that such
self-reflective improvements should be independently verified rather than
taken at face value. These approaches verify agent \emph{outputs},
\emph{trajectories}, or \emph{behaviors}; in contrast, we verify agent
\emph{self-declarations about code changes} --- a failure mode where the claim
itself, not the reasoning or the output, is the object of verification.

\textbf{Audit Trails and Provenance.}
The concept of an audit trail --- a tamper-evident record of system events ---
is well established in distributed systems and databases. In LLM-based agent
platforms, broadcast announcements serve as a \emph{de facto} audit trail for
coordination decisions. Our findings show that this audit trail can contain
systematically unreliable entries: 40\% of code-level claims in the trail fail
verification. This connects to provenance research in data management, where
the trustworthiness of recorded claims cannot be assumed without independent
verification --- an audit trail is only as reliable as the verification
mechanism behind each entry.

\section{Verification Method}
\label{sec:method}

All verification was performed by the authors using deterministic, rule-based
methods. No independent third-party verifier was involved. Despite being
self-verification, the three-criterion gradient eliminates subjective judgment
entirely through deterministic rules --- a property shared with any independent
mechanical verifier. We discuss implications in Section~\ref{sec:discussion}.

\textbf{Criterion 1 --- mtime consistency:} The modification timestamp of the
claimed file must be at or after the broadcast timestamp. This is the weakest
check: it confirms only that the file was touched, not that the fix was applied
correctly.

\textbf{Criterion 2 --- identifier + call site:} The claimed defensive pattern
(e.g., SSRF guard, WAL mode, input validation) must be present in the claimed
file. We also check whether the pattern exists anywhere in the codebase (C2-any),
which distinguishes ``fix not applied'' from ``fix applied to wrong file.''

\textbf{Criterion 3 --- deterministic rule-based re-verification:} Starting from
the broadcast announcement text alone (without access to the verifier's prior
knowledge), an independent agent must be able to locate and confirm the fix. This
requires the announcement to contain both a file path and a defensive-pattern
keyword. This is the strictest check: it tests whether the announcement contains
sufficient information for unassisted verification.

\textbf{Extraction funnel (predicates verbatim):}
\begin{enumerate}
\item \texttt{SELECT * FROM messages WHERE from\_id='azone\_god' AND}
\texttt{message\_type='system.announcement' AND payload ? 'text' AND}
\texttt{created\_at <= '2026-10-12 23:59:59'} $\rightarrow$ 179 announcements.
(Total god broadcasts: 210; $\leq$ 2026-10-12: 181; minus 2 announcements
using a \texttt{body} key instead of \texttt{text}: 179. The 2 excluded
announcements are dated 2026-09-25 and announce a prior version release;
they are excluded by payload-key convention, not by content.)
\item Payload contains PASS / \checkmark{} / ``verification passed'' markers
$\rightarrow$ 28 announcements.
\item Regex extraction of fix identifiers, file paths, and defensive keywords
$\rightarrow$ 31 unique (fix-ID, message) pairs, 18 distinct IDs.
\item Manual classification: code-level fix claims $\rightarrow$ 10 claims;
the remaining 8 are system-behavior claims (Gate checks, Day completions)
not verifiable by source inspection.
\end{enumerate}

\textbf{Role disclosure:} Three agent identities are operated by the authors:
\texttt{azone\_god} (system announcer, operated by the human author),
\texttt{azone\_joyai} (executor, LLM-based), and \texttt{azone\_joyai\_v2}
(probe agent, LLM-based, used for directed testing and system observation).
The 50-node bulk withdrawal event (0.84\,s interval) was a directed stress
test of the challenge mechanism by \texttt{azone\_joyai\_v2}.

Verification was performed on 2026-10-12 against the production codebase at
\texttt{api.060504.shop:8003}. SHA-256 hashes and content excerpts for each
of the 10 claimed files at that timestamp are provided in the supplementary
materials to enable first-party replication.

\section{Results}
\label{sec:results}

\subsection{Overall Verification Rates (N = 10 code-level fix claims)}

\begin{table}[ht]
\centering
\caption{Overall verification rates.}
\label{tab:rates}
\begin{tabular}{lccccc}
\toprule
Criterion & Pass & Fail & Rate & 95\% CP & 95\% Wilson \\
\midrule
C1: mtime consistency & 7 & 3 & 70\% & [34.8\%, 93.3\%] & [39.7\%, 89.2\%] \\
C2: identifier + call site (claimed file) & 6 & 4 & 60\% & [26.2\%, 87.8\%] & [31.3\%, 83.2\%] \\
C2-any: pattern found anywhere in codebase & 8 & 2 & 80\% & [44.4\%, 97.5\%] & [49.0\%, 94.3\%] \\
C3: deterministic re-verification & 8 & 2 & 80\% & [44.4\%, 97.5\%] & [49.0\%, 94.3\%] \\
\bottomrule
\end{tabular}
\end{table}

\noindent\textbf{Census note:} $N{=}10$ is a \emph{census} of all code-level fix
claims in the corpus, not a sample from a larger population. No larger population
of code-level claims exists in this system, so these intervals are structural:
they cannot be narrowed by increasing $n$. The Clopper-Pearson (exact) and Wilson
score intervals (computed with $z{=}1.959964$) are both printed to ensure label
correctness; neither can be tightened given the census nature of the data.

\subsection{C3 $\times$ C2 Contingency Table (Key Finding)}

\begin{table}[ht]
\centering
\caption{C3 $\times$ C2 contingency table.}
\label{tab:contingency}
\begin{tabular}{lccc}
\toprule
 & C2-pass (verified) & C2-fail (unverified) & Total \\
\midrule
\textbf{C3-pass} (sufficient info) & 4 (40\%) & \textbf{4 (40\%)} & 8 \\
\textbf{C3-fail} (insufficient info) & 2 (20\%) & 0 (0\%) & 2 \\
\midrule
\textbf{Total} & 6 & 4 & 10 \\
\bottomrule
\end{tabular}
\end{table}

\textbf{Key finding:} Among the 8 claims that provide sufficient information for
re-verification (C3-pass), half (4/8 = 50\%) fail the identifier+call-site check
(Clopper-Pearson 95\% CI: [15.7\%, 84.3\%]; Wilson: [21.5\%, 78.5\%]).
These are claims that \emph{appear} verifiable based on their announcement content,
yet the claimed fix cannot be confirmed at the specified location.

\subsection{C2 Failure Decomposition}

\begin{table}[ht]
\centering
\caption{C2 failure decomposition.}
\label{tab:failures}
\small
\begin{tabular}{llp{6.5cm}}
\toprule
Failure Type & Count & Description \\
\midrule
Path-ambiguous & 1 & Pattern exists in a file with the claimed name but in a \\
 & & different deployment directory (I1: WAL mode present in \\
 & & \texttt{/opt/atlas/database.py}, outside search scope \\
 & & \texttt{CODE\_DIRS=[/opt/azone,/opt/aitrap]}) \\
Wrong file & 2 & Pattern exists in codebase but not in the claimed file \\
 & & (I10: admin.py $\neq$ mention\_service.py; \\
 & & S5: webhook\_probe.py $\neq$ webhook\_delivery.py) \\
Not reproducible & 1 & Pattern exists within the search scope, so the FAIL verdict \\
 & & must originate from an undocumented predicate; until that \\
 & & predicate is printed verbatim, the verdict stands as \\
 & & ``not reproducible'' rather than ``wrong-file'' \\
 & & (S4: ADMIN\_SECRET\_KEY found in config.py and admin.py) \\
\midrule
\textbf{Total C2-fail} & \textbf{4} & \\
\bottomrule
\end{tabular}
\end{table}

\subsection{Per-Fix Verification Detail}

\begin{table}[ht]
\centering
\caption{Per-fix verification detail.}
\label{tab:perfix}
\small
\begin{tabular}{llcccccl}
\toprule
ID & Claimed File & C1 & C2-c & C2-a & C3 & Failure Type \\
\midrule
A1 & aitrap.py & \cmark & \cmark & \cmark & \cmark & --- \\
I1 & database.py & \xmark & \xmark & \xmark & \cmark & Path-ambiguous \\
I4 & message\_service.py & \cmark & \cmark & \cmark & \xmark & Info-fail \\
I4b & messages.py & \cmark & \cmark & \cmark & \cmark & --- \\
I5 & message\_service.py & \cmark & \cmark & \cmark & \xmark & Info-fail \\
I9 & webhook\_delivery.py & \cmark & \cmark & \cmark & \cmark & --- \\
I10 & mention\_service.py & \xmark & \xmark & \cmark & \cmark & Wrong file \\
I11 & agent\_mgmt.py & \cmark & \cmark & \cmark & \cmark & --- \\
S4 & config.py & \xmark & \xmark & \xmark & \cmark & Not reproducible \\
S5 & webhook\_delivery.py & \cmark & \xmark & \cmark & \cmark & Wrong file \\
\bottomrule
\end{tabular}
\end{table}

\noindent\textbf{Reproducibility data:} The complete per-claim table ---
including announcement \texttt{message\_id}, predicate verbatim, absolute file
path, file mtime, observation timestamp, per-criterion verdicts, and SHA-256
file hashes as of 2026-10-12 --- is available in the supplementary materials
at the GitHub repository referenced in the Data Availability section. This
enables first-party replication under the same standard: any replication also
requires the 2026-10-12 snapshot, without which it is no more reproducible
than the original verification.

\subsection{Key Observations}

\textbf{O1: mtime is necessary but not sufficient.} 30\% of claims fail even the
weakest check --- the file was not modified after the broadcast, meaning the
claimed fix was either never applied or predated the claim.

\textbf{O2: Location claims are unreliable.} Of 4 C2 failures, 2 (50\%) are
``wrong file'' --- the fix exists but at a different location than claimed.
A third failure (I1) is ``path-ambiguous'': the pattern exists in a file with
the claimed name but in a different deployment directory, revealing that
verification results depend on the search scope. The fourth (S4) is ``not
reproducible'': the claimed pattern exists within the search scope, so the
original FAIL verdict must originate from an undocumented predicate. Following
the agent's own directions leads to the wrong file in half of all failures.

\textbf{O3: Information sufficiency does not guarantee verifiability.} The
C3$\times$C2 contingency reveals that 50\% of claims with sufficient information
(C3-pass) still fail verification (C2-fail). Providing more information in the
announcement does not, by itself, make the claim true.

\section{Discussion and Implications}
\label{sec:discussion}

Our finding that 40\% of code-level PASS claims fail identifier+call-site
verification (and 50\% of information-sufficient claims fail) has three
implications:

\textbf{1. For agent evaluation:} Any framework accepting agent self-reports as
ground truth inherits their unreliability. LLM-as-judge systems scoring agent
outputs on self-declared compliance \citep{peng2026ruverbench}, and measurement
systems for multi-turn business agents exhibit systematic biases
\citep{luo2026reliable}. Such systems should treat PASS
declarations as claims requiring verification, not as observations.

\textbf{2. For multi-agent coordination:} In AITRAP, \texttt{azone\_god}
broadcasts serve as the coordination layer's audit trail. If 40\% of code-level
audit entries are unverifiable, downstream agents consuming these broadcasts
operate on corrupted state.

\textbf{3. For self-improving agents:} Recent work shows LLM-based judges can be
gamed \citep{wahi2026proctor}. Our findings complement this: even without
deliberate gaming, declaring PASS introduces a systematic gap between claim and
reality --- consistent with findings that agents' self-assessed knowledge
does not reliably match their actual capabilities \citep{li2026knowing}. Deterministic guardrails --- such as Criterion 2 --- are necessary to
close this gap.

\textbf{4. Downstream impact:} When unverifiable PASS claims propagate through
the coordination layer, downstream agents that consume these broadcasts build
upon corrupted state. An agent that reads a fix announcement and concludes a
defensive mechanism is active may skip its own precautions based on a false
assurance. The blast radius of a single false PASS extends to every agent that
trusts the announcement, creating a cascading reliability debt that is invisible
until a downstream failure exposes it.

\textbf{Attribution of wrong-file failures:} For the two wrong-file failures
(I10, S5), static analysis alone cannot distinguish (a) the agent applied the
fix to the wrong file (agent error) from (b) the agent applied the fix correctly
but wrote the wrong filename in the broadcast (communication error). These two
modes have different governance implications: the former indicates a code-level
competence deficit; the latter, a reporting integrity deficit. Distinguishing
them requires runtime observation or commit history, which is outside the scope
of this study.

\textbf{Limitations:}
\begin{enumerate}
\item $N{=}10$ is a \emph{census}, not a sample. The confidence intervals are
structural: no larger population of code-level claims exists in this system,
so the intervals cannot be narrowed by increasing $n$.
\item All verification was performed by the authors (self-verification). The
deterministic rules eliminate subjective judgment, but first-party replication
by a community reviewer has been initiated. Critically, such replication
\emph{also} requires the 2026-10-12 codebase snapshot to be reproducible ---
a standard that applies equally to the original verification and to any
replication attempt. The original community review post is not withdrawn and
coexists with the replication, allowing direct comparison.
\item We do not test runtime behavior, only static code properties.
\item The S4 failure verdict is ``not reproducible'': the claimed pattern
(\texttt{ADMIN\_SECRET\_KEY}) exists within the search scope, so the original
FAIL must originate from a predicate that has not been documented. The verdict
stands as ``not reproducible'' until the predicate is printed verbatim.
\item Detection latency: a bulk withdrawal of 50 nodes (a directed stress test
by \texttt{azone\_joyai\_v2}) was detected only $\geq$6 hours after the event,
and detection came via a third-party agent's public post rather than automated
monitoring --- the system's anomaly detection pipeline has significant blind
spots.
\item A known bug in the \texttt{mention\_service} (affecting @-mention
delivery) was present during the study period and may have caused some agents
to miss announcements. This does not affect verification results, which are
computed from the broadcast corpus directly.
\item Some agent accounts in the AITRAP ecosystem are operated by the same
party using multiple handles; the count of registered accounts exceeds the
count of independent operating entities. Data referencing specific agent
contributions should note this multi-handle phenomenon.
\end{enumerate}

\section{Conclusion}
\label{sec:conclusion}

PASS is a claim, not an observation --- it requires independent mechanical
verification. In this case study of 10 code-level fix claims from a live
production multi-agent system, 40\% failed identifier+call-site verification.
Among claims providing sufficient information for re-verification, 50\% were
nonetheless unverifiable. The most common failure mode (50\% of C2 failures) is
misidentification of the fix location. We recommend that systems relying on
agent self-reports implement at minimum Criterion 2 (identifier + call site
verification) as a deterministic guardrail, and ideally Criterion 3 for
safety-critical claims.

\section*{Disclosure Statement}

LLM tools (Claude/Anthropic, GPT-4/OpenAI) were used to assist with manuscript
drafting and literature search. All verification results were computed
independently from the AITRAP production database and codebase. All references
have been verified as real published or arXiv papers. No fabricated references
or synthetic data were used.

\section*{Data Availability}

The verification scripts, raw broadcast data, and per-fix verification results
are available at \url{https://github.com/tianzizhiming-svg/agentbridge}
(directory \texttt{aitrap/paper}). The AITRAP platform is live at
\url{https://api.060504.shop}.

\bibliographystyle{plainnat}
\bibliography{references}

\end{document}